# Analysis of Radiation Level and Estimation of Protection Distance of γ Mobile Flaw Detection Source"

Liu Zhihui[1]    Hu Xiang[1]    Zhang Zhengyang[2]

1. Beijing Nuclear and Radiation Safety Center, Beijing, China

2. University of Cambridge, UK

**Abstract:** **[Objective]** To analyze the radiation dose associated with gamma-ray mobile flaw detection, estimate the extent of the supervision and control areas, and assess the associated radiation risks. **[Methods]** A combination of theoretical calculations and actual measurements was used to compare and analyze the ambient equivalent dose rates of $^{192}$Ir and $^{75}$Se at their nominal source strengths. Measurements were conducted at distances of 1 m, 2 m, and 5 m from the radiation source. The extents of the control and supervision areas were estimated under three working scenarios: (1) without considering air attenuation, (2) considering air attenuation, and (3) after shielding by the flaw detection workpiece, using source activities of $3.7 \times 10^{10}$ Bq and $3.7 \times 10^{12}$ Bq. **[Results]** Actual measurement of radiation dose of $^{192}$Ir and $^{75}$Se were measured under three different nominal activities. Theoretical calculation of radiation dose estimates at various distances were obtained for both nuclides, and the results showed that the theoretical values were basically consistent with the measured values. **[Conclusion]** The estimated scope of the supervision and control areas provided in this study can serve as a reference for flaw detection companies. Technicians can use these estimates to calculate appropriate distances for safety zones based on different nuclide activities. This enables flaw detection personnel to reduce the measurement scope on-site and to quickly and accurately define area boundaries.

Risk industrial γ flaw detection creates certain economic benefits, it also has inherent risks, storage risks, transportation risks, operational risks, and decommissioning or source replacement risks [1]. If field operations and management are improper, radiation accidents may occur. Of the 17 radiation accidents that occurred from 2004 to 2013, 15 were directly or indirectly related to mobile γ flaw detection [2]. In order to illustrate the potential dangers of γ flaw detection sources, this paper estimates the dose equivalent rates generated by two radionuclides ($^{192}$Ir and $^{75}$Se) and three nominal activity radioactive sources, and compares and analyzes them with measured data, calculates the ambient equivalent dose rate, and estimates the scope of the control area and the supervision area, objectively characterizing the potential radiation hazards of γ flaw detection to occupational personnel.

1. Objects and methods

1.1 Research subjects

$^{192}$Ir and $^{75}$Se radioactive sources in a source library as an example, the radiation level around the radioactive source is analyzed. When the thickness of the steel workpiece is 20mm, the range of the supervision area and the control area is given.

1.2 Instruments and methods

1.2.1 Testing instruments

were used : FH40G portable X-γ dose rate meter (range: 100nSv/h-1Sv/h) from Therm, USA, and identiFINDER portable gamma spectrometer (range: 0.01μSv/h-1Sv/h) from Target, Germany. Both instruments were calibrated by the China Institute of Metrology and were within the validity period.

1.2.2 On-site testing

Based on the "Radiation Health Protection Standard for Sealed Radioactive Sources and Sealed Gamma Radiation Source Containers" GBZ 114-2006; "Radiation Safety Requirements for Gamma Ray Flaw Detection Devices", Environmental Protection (2007) No. 8; "Dose Conversion Factors for Photon External Exposure Radiation Protection" (GBZ/T144-2002) [3].

This measurement: In a source library flaw detection experimental site, the mobile flaw detection device containing radioactive sources $^{192}$Ir and $^{75}$Se was measured for radiation levels at three different activities of $4.81\times10^8$ Bq, $3.7\times10^9$ Bq, and $3.7\times10^{10}$ Bq for $^{192}$Ir and $1.48\times10^9$ Bq, $3.7\times10^9$ Bq, and $3.7\times10^{10}$ Bq for $^{75}$Se. The radiation sources of each activity of each nuclide were measured at 1 meter, 2 meters, and 5 meters.

1.2.3 Theoretical estimation

1.2.3.1 Relationship between kerma and absorbed dose

3.3.7 of Ionizing Radiation Protection and Radiation Source Safety (edited by Pan Ziqiang and Cheng Jianping) [4], since the energy of the gamma rays emitted by nuclides $^{192}$Ir and $^{75}$Se is not high, the bremsstrahlung radiation of the secondary charged particles produced can be ignored, and the kerma and absorbed dose are equal. The following estimated results of absorbed dose are equivalent to the kerma.

1.2.3.2 Estimation of absorbed dose

2.5.5.2 and Table 2.11 of the Radiation Protection Handbook – Volume 3: Radiation Safety [5], the following formula (1) is used to estimate the air absorbed dose rate at 1 m, 2 m and 5 m respectively.

$$\dot{X} = 0.235\frac{A\Gamma}{r^2} \qquad (1)$$

Where $\dot{X}$: air absorbed dose rate (μGy/h); A: source activity (MBq); r: distance from measurement position to source (m); Γ: irradiation

rate constant (R m$^2$/ (hCi) ). Γ$_{Ir-192}$ = 0.497 Rm$^2$/ (hCi) , Γ$_{Se-75}$ = 0.204 Rm$^2$/ (hCi )

1.2.3.3 Relationship between kerma, ambient dose equivalent and effective dose

Section 4.1a of the Dose Conversion Coefficients for Photon External Irradiation Protection (GBZ/T144-2002) [3], the ambient dose equivalent H*(10 ) at a point in space can be used as an approximate value of the effective dose to the human body at that point. Therefore, the following estimation results of H*(10 ) can be used as approximate values of the effective dose.

Section 6.2a of the Dose Conversion Coefficients for External Photon Radiation Protection (GBZ/T144-2002) [3] and Table B$_1$ in Appendix B, the estimated results of absorbed dose/kerma are converted to H*(10).

1.2.3.4 Estimation of ambient dose equivalent

B1 in Appendix B of "Dose Conversion Factors for Photon External Irradiation Radiation Protection" (GBZ/T144-2002) [3] only gives the conversion factor between the kerma and ambient dose equivalent H*(10 ) of monoenergetic gamma rays , it is also necessary to weight the conversion factor of gamma rays of different energies emitted by nuclides according to the gamma ray intensity, and then convert the kerma into the ambient dose equivalent H*(10 ) . The values of the gamma ray intensity emitted by $^{192}$Ir and $^{75}$Se are taken from "Table of Isotope " [6], ignoring gamma rays with an intensity less than 1%.

1.2.3.5 Preliminary estimation of the distance between the surveillance area, control area boundary and source

7.3.1 and 7.3.6 of the Industrial Gamma Ray Flaw Detection Protection Standard (GBZ132-2008) , for on-site flaw detection operations, the air kerma rate outside the control area boundary should be lower than 1.5μGy/h , and the air kerma rate at the supervision area boundary should not exceed 2.5μGy/h . At the same time, refer to Appendix C of the standard to estimate the distance between the control area and the supervision area.

1.2.3.6 Attenuation of the γ-ray kerma of a radioactive source by air

Section 5.2 and Table 5.7 of the Radiation Protection Handbook – Volume 1 : Radiation Sources and Shielding [7], the attenuation of the γ-ray kerma of the radioactive source by air is estimated using formulas (2), (3), and (4) [8]. The air density is set to 1.29 g/L , and the composition is set to $N_2$: 78% , $O_2$:21% , and Ar: 1% by volume.

$$K_0 = \frac{A * \Gamma_k}{R^2}; \quad (2)$$

$$K = K_0 * e^{-\mu*d}; \quad (3)$$

$$K = \frac{A * \Gamma_k}{R^2} * e^{-\mu*d} \quad (4)$$

Where, K0 is the air absorbed dose rate before shielding, K is the air absorbed dose rate after shielding, A: source activity (MBq); R : distance from the measurement position to the source (m); $\Gamma_k$ : irradiation rate constant ( $Rm^2$ /(hCi)). μ is the γ linear attenuation coefficient of the shielding material.

$\Gamma_{Ir-192}$=3.18 Gy·$m^2$/Bq·s , $\Gamma_{Se-75}$=4.25E-17Gy·$m^2$/Bq·s [9], after dose conversion $\Gamma_{Ir-192}$=0.13 mGy·$m^2$·$h^{-1}$·$GBq^{-1}$, $\Gamma_{Se-75}$=0.265 mGy·$m^2$·$h^{-1}$·$GBq^{-1}$. [8]

2. Results

2.1 Theoretical estimation

The detection thickness range of the flaw detection workpiece is 20～100mm for $^{192}$Ir and $^{75}$Se The theoretical estimation is based on the estimated ambient dose equivalent rate at 1m , 2m and 5m, with $^{192}$Ir at nominal 4.81×$10^8$Bq, 3.7×$10^9$Bq, and 3.7×$10^{10}$Bq, and $^{75}$Se at nominal 1.48×$10^9$Bq, 3.7×$10^9$Bq, and 3.7×$10^{10}$Bq, respectively. The estimated results are shown in Table 1.

Table 1 Theoretical estimated values of dose rates at different distances for different nominal activities of $^{192}$Ir and $^{75}$Se

| | $^{192}$Ir ( Bq ) | | | $^{75}$Se ( Bq ) | | |
|---|---|---|---|---|---|---|
| | $4.8\times10^{8}$ | $3.7\times10^{9}$ | $3.7\times10^{10}$ | $1.48\times10^{9}$ | $3.7*10^{9}$ | $3.7\times10^{10}$ |
| distance | Dose rate ( μGy/h ) | | | Dose rate ( μGy/h ) | | |
| 1m | 47.45 | 412.15 | 4121.50 | 69.86 | 174.64 | 1746.40 |
| 2m | 11.86 | 103.04 | 1030.38 | 17.46 | 43.66 | 436.60 |
| 5m | 1.90 | 16.49 | 164.86 | 2.79 | 6.99 | 69.86 |

2.2 Test results

In the gamma flaw detection source library test field, two portable dose rate meters, FH40G and identiFINDER, were used to measure the equivalent dose rate around the source. The activity of the source was $^{192}$Ir: $4.81\times10^{8}$ Bq, $3.7\times10^{9}$ Bq and $3.7\times10^{10}$ Bq; $^{75}$Se: $1.48\times10^{9}$ Bq, $3.7\times10^{9}$ Bq and $3.7\times10^{10}$ Bq. The measurement results of FH40G (after dose conversion and considering the calibration factor) are shown in Table 2:

Table 2 Actual measured values of three different nominal activities of $^{192}$Ir and $^{75}$Se

| | $^{192}$Ir ( Bq ) | | | $^{75}$Se ( Bq ) | | |
|---|---|---|---|---|---|---|
| | $4.8\times10^{8}$ | $3.7\times10^{9}$ | $3.7\times10^{10}$ | $1.48\times10^{9}$ | $3.7\times10^{9}$ | $3.7\times10^{10}$ |
| distance | Dose rate ( μGy/h ) | | | Dose rate ( μGy/h ) | | |
| 1m | 14.23 | 351.00 | 2642.08 | 47.26 | 120.18 | 1009.18 |
| 2m | 4.10 | 97.64 | 817.71 | 12.86 | 35.58 | 297.52 |
| 5m | 0.79 | 15.80 | 148.62 | 2.68 | 8.58 | 54.84 |

using the identiFINDER portable instrument are shown in Table 3:

Table 3 Actual measured values of three different nominal activities of $^{192}$Ir and $^{75}$Se

| | $^{192}$Ir ( Bq ) | | | $^{75}$Se ( Bq ) | | |
|---|---|---|---|---|---|---|
| | $4.81\times10^{8}$ | $3.7\times10^{9}$ | $3.7\times10^{10}$ | $1.48\times10^{9}$ | $3.7\times10^{9}$ | $3.7\times10^{10}$ |
| distance | Dose rate ( μGy/h ) | | | Dose rate ( μGy/h ) | | |
| 1m | 15.79 | 372.33 | 3031.14 | 44.71 | 160.67 | 1506.67 |
| 2m | 4.97 | 91.17 | 1103.36 | 15.67 | 37.00 | 544.67 |

| 5m | 0.92 | 18.75 | 217.10 | 3.20 | 8.61 | 54.60 |

### 2.3 Comparison of measured results and theoretical estimates

Comparison of the theoretical estimated value and actual measured value of $^{192}$Ir of mobile gamma flaw detection source: When the activity of $^{192}$Ir is $4.81\times10^{8}$ Bq, the activity is too low and is greatly affected by the surrounding scattered rays. It is very different from the theoretical value and not representative. This set of data is excluded and not compared. The comparison of the other two sets of data ($^{192}$Ir at $3.7\times10^{9}$ Bq and $3.7\times10^{10}$ Bq) is shown in Figure 1.

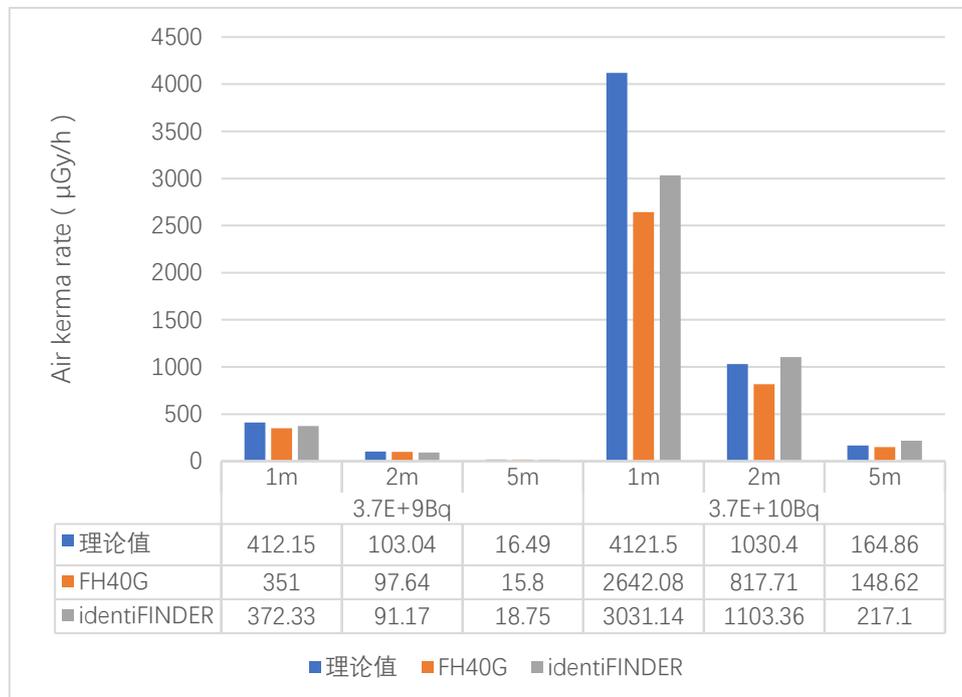

Fig. 1 Comparison between the estimated values of $^{192}$Ir activity $3.7\times10^{9}$ Bq and $3.7\times10^{10}$ Bq and the field measurement results

Comparison between the theoretical estimated value and the actual measured value of $^{75}$Se by mobile γ flaw detection source: The comparison of three groups of data of $^{75}$Se ($1.48\times10^{9}$ Bq, $3.7\times10^{9}$ Bq, $3.7\times10^{10}$ Bq) is shown in Figure 2.

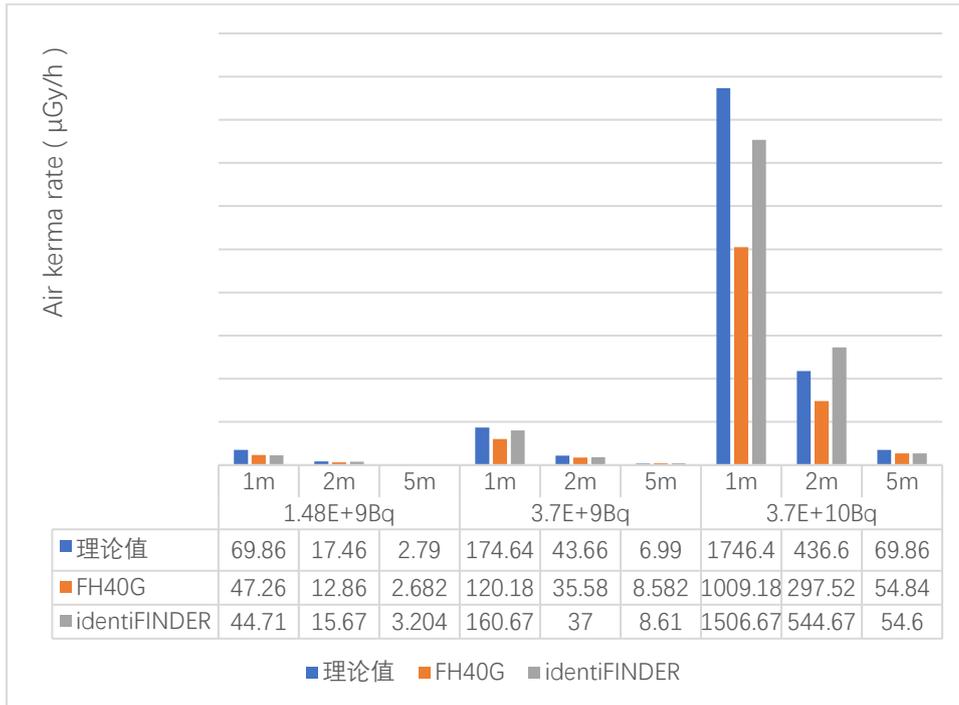

Figure 2 $^{75}$Se 1.48×10$^9$Bq, 3.7×10$^9$Bq, 3.7×10$^{10}$Bq comparison of estimated values with field measurements

2.4 Protection distance of supervision area and control area

The dose rate at the boundary of the supervision area is 2.5 μGy/h, while the dose rate at the boundary of the control area is 15 μGy/h. When the source activity is 3.7 × 10$^{10}$ Bq (which corresponds to the maximum activity measured in this study) or 3.7 × 10$^{12}$ Bq, the estimated protection distances for the mobile γ-ray flaw detection source—under three conditions: without considering air scattering, with air scattering considered, and with shielding provided by the tested workpiece—are shown in Table 4.

Table 4 Boundaries of control and supervision areas under different shielding conditions 3.7×10$^{10}$ Bq and 3.7×10$^{12}$ Bq

| shield Condition | use Nuclide | activity (Bq) | thickness (mm) | Control area boundary (m) | Supervision area boundary (m) |
|---|---|---|---|---|---|
| No shield | Ignore air scattering | $^{75}$Se | 3.7E×10$^{10}$ | -- | 11.6 | 28.2 |
| | | $^{192}$Ir | 3.7E×10$^{10}$ | -- | 18.0 | 43.8 |
| | | $^{75}$Se | 3.7E×10$^{12}$ | -- | 116 | 282 |
| | | $^{192}$Ir | 3.7E×10$^{12}$ | -- | 179 | 438 |
| in | Consider | $^{75}$Se | 3.7E×10$^{10}$ | -- | 10.2 | 21.8 |

| | | | | | |
|---|---|---|---|---|---|
| g air scattering * | $^{192}$Ir | 3.7E×10$^{10}$ | -- | 14.3 | 30.3 |
| | $^{75}$Se | 3.7E×10$^{12}$ | - | 102 | 218 |
| | $^{192}$Ir | 3.7E×10$^{12}$ | - | 143 | 302 |
| Workpiece shielding ** | $^{75}$Se | 3.7E×10$^{10}$ | 20 | 6.4 | 15.6 |
| | $^{192}$Ir | 3.7E×10$^{10}$ | 20 | 10.2 | 25.0 |
| | $^{75}$Se | 3.7E×10$^{12}$ | 20 | 64 | 156 |
| | $^{192}$Ir | 3.7E×10$^{12}$ | 20 | 102 | 250 |

Air scattering * $K_0 = \frac{A*\Gamma_k}{R^2} * e^{-u_{en}*d}$, workpiece shielding ** $F = \frac{3.6 \times 10^6 \times \Gamma \times A}{R^2 \times \alpha^{T/t_0}}$

3 Discussions

To minimize discrepancies between instruments [10]-[11], two different devices were used simultaneously during measurements. This approach helped reduce systematic deviations caused by instrument-related errors. When measuring on site, the angular response of the instrument to the radiation field [12], the energy response [13], and the instrument range overload [14] should be considered. If the flaw detector is required to calculate the distance between the supervision area and the control area on site, not only will the calculation be cumbersome, but also due to the uneven technical quality of the flaw detectors, most of the front-line flaw detectors are not able to calculate the protective distance of the supervision area and the control area. The front-line flaw detectors only use instrument measurements to find the boundaries of the supervision area and the control area. It takes a long time, and when the boundaries are drawn blindly, the exposure dose of the operators is increased, and the boundaries are basically not drawn. This method is time-consuming, and when boundaries are drawn arbitrarily or inaccurately, it can lead to increased radiation exposure for operators—and in many cases, the boundaries are not marked at all.

When high-activity sources are used in the field, user units typically employ collimators or other protective devices and methods to ensure safety. However, when dealing with the source strength becoming small, enterprises often overlook necessary radiation protection measures. This study specifically focuses on low-activity sources that have not been decommissioned and are still in use for inspecting thin workpieces, conducting both theoretical calculations and actual measurements.

It is recommended that the flaw detection company arrange full-time technical personnel in advance to regularly calculate the nominal activity of each source according to factors such as half-life, and make an EXCEL table based on the distance between the supervision zone and the control zone in the existing public data [8][15], so as to estimate the approximate range of the supervision zone and the control zone in advance. Train the flaw detection personnel so that they can know the range of the control zone and the supervision zone at a glance. Combined with actual measurements on site, the demarcation time can be greatly reduced and the exposure dose during demarcation can be greatly reduced. Supervision areas are set up to reduce the dose to NDT personnel, and control areas are set up to prevent the public from being exposed. At the same time, it prevents unscrupulous people who want to make trouble from breaking in and blackmailing from the public.

As shown in Table 2, the distances between the supervision and control zones vary significantly depending on the source activity. Since mobile flaw detection is often influenced by site conditions such as terrain, limited space, and nearby shielding objects that may obstruct visibility, it is recommended that companies assign dedicated technical personnel to regularly assess the nominal activity of each source, taking into account factors such as radioactive half-life.

Flaw detection personnel should be trained to recognize these distances at a glance. By combining pre-estimated data with on-site measurements, the time required to delineate safety zones can be significantly reduced, along with the radiation dose received during this process. Supervision zones are intended to minimize exposure for NDT personnel, while control zones are established to protect the public. These zones also serve to prevent unauthorized individuals from entering the area, which could lead to public disturbances or extortion attempts.

NDT activities, especially field NDT, cannot rely solely on regulatory authorities. Users must establish a sense of subjectivity and strengthen their own management. Regulatory authorities should conduct random inspections on the quality of NDT machines before they leave the factory to ensure that the measurement and protection meet the standards before they leave the factory. They should also urge NDT companies to fully implement all radiation safety and protection measures [16]. NDT companies should purchase NDT machines from regular NDT production units, control the quality of NDT equipment before they leave the factory, ensure that the source strength is accurate and the radiation protection meets the standards. They should also formulate and conduct targeted emergency plans for radiation accidents. In addition, in terms of operation, it should be ensured that the NDT machines are regularly inspected and maintained, that the records of entry and exit of the source warehouse are kept, that the operators strictly abide by the operating procedures during NDT, and that they protect themselves well to avoid accidental exposure [17].

In short, we should actively cultivate radiation safety culture and regard safety culture as part of defense in depth [18]-[19]. Only by comprehensively managing the entire process from supervision, production, use, transportation, and decommissioning can we minimize the occurrence of radiation accidents.